\documentclass[AMA,STIX1COL]{WileyNJD-v2}
\usepackage[utf8]{inputenc}
\def\doiheadtext{\relax}
\usepackage{underscore}
\usepackage{listings}

\lstdefinestyle{customc}{%
  belowcaptionskip=1\baselineskip,
  breaklines=true,
  xleftmargin=\parindent,
  language=C,
  showstringspaces=false,
  basicstyle=\ttfamily\normalsize,
  keywordstyle=\bfseries\color{green!30!black},
  numberstyle=\tiny,
  commentstyle=\color{purple!30!black},
  identifierstyle=\bfseries\color{black},
  stringstyle=\color{orange},  morekeywords={uint64_t,uint32_t,int64_t,__int128_t,int32_t,__m256i,__m512i,__m128i,UINT64_C},
}
\newcommand{\noop}[1]{} 

\usepackage{subfloat,subfig}
\usepackage{siunitx}
\date{January 2019}
\usepackage{url}

\usepackage{underscore}

\usepackage{tikz}
\usetikzlibrary{chains,arrows,automata,decorations.markings,positioning,calc,decorations.pathreplacing,patterns}

\articletype{Short Communication}%

\makeatother

\begin{document}

\title{Base64 encoding and decoding at almost the speed of a memory copy}

\author[1]{Wojciech Muła}

\author[1]{Daniel Lemire*}

\authormark{Wojciech Muła and Daniel Lemire}
\address[1]{\orgdiv{TELUQ}, \orgname{Universit\'e du Qu\'ebec}, \orgaddress{\state{Quebec}, \country{Canada}}}

\corres{*D. Lemire, Universit\'e du Qu\'ebec (TELUQ), 5800,  Saint-Denis street, Montreal (Quebec)  H2S 3L5, Canada. \email{daniel.lemire@teluq.ca}}


\abstract[Summary]{
Many common document formats on the Internet  are text-only
such as email (MIME) and the Web 
(HTML, JavaScript, JSON and XML). To include 
images or executable code in these documents, we first encode them as text using base64.
Standard base64 encoding uses 64~ASCII characters: both lower and upper case Latin letters, digits and two other symbols.
We show how we can encode and decode base64 data at nearly the speed of a memory copy (\texttt{memcpy}) on recent Intel processors, as long as the data does not fit in the first-level (L1) cache. 
We use the SIMD (Single Instruction Multiple Data) instruction set AVX-512 available on commodity processors.
Our implementation generates several times fewer instructions than previous SIMD-accelerated base64 codecs. It is also more versatile, 
as it can be adapted---even at runtime---to any base64 variant by only changing constants.




}



\maketitle

\section{Introduction}\thispagestyle{empty}

Base64 formats represent arbitrary binary data as ASCII text: e.g., the ubiquitous MIME email protocol\cite{rfc2045} require that arbitrary binary attachments be encoded as ASCII, typically using base64. In a similar manner, binary data (e.g., images, WebAssembly code) can also be included inside text-only resources like XML, JavaScript or HTML documents. Base64 is part of the standard library of many popular programming languages (Java, Go, JavaScript, Swift), it is used in many important database systems. Some database systems even store binary data as base64 (MongoDB, Elasticsearch, Amazon SimpleDB and DynamoDB). Crane and Line\cite{Crane:2017:ESA:3121050.3121086} observe that decoding all this base64 data can be a performance concern.


Commodity processors support single-instruction-multiple-data (SIMD) instructions. 
The SIMD extensions AVX-512 available on recent Intel processor operate on 512-bit registers.
Hence, we can compare two strings of 64~characters using a single instruction.
Software that uses fewer instructions\cite{Cebrian2019} tend 
to be faster and use less energy.

We refer to algorithms designed to benefit
from SIMD instructions as being vectorized.
We can reasonably expect that as new processor
architectures include wider SIMD registers, corresponding vectorized algorithms can run 
faster. A doubling of the register width 
might almost double the performance---everything else being equal.
In earlier\cite{simdbase64} work, we showed
that we could 
greatly increase base64 encoding and decoding speeds
by using 256-bit registers available in AVX and AVX2 instruction set extensions.
Such a vectorized approach was later adopted by the PHP interpreter and other systems.

Maybe surprisingly, we find that we can more than double the speed by using the new 512-bit instructions: for example, we achieve a five-fold reduction in the number of instructions during decoding by moving from AVX2 to AVX-512. For inputs that do not fit in the fastest CPU cache (L1), it is nearly exactly as fast to copy base64 data than to encode or decode it.
The key to this greater-
than-expected performance is an algorithmic redesign to benefit from the new instructions introduced in AVX-512.

\section{Base64}
\label{sec:base64}

The ASCII character set is made of 128~code points with byte values in $[0,128)$; we can recognize a stream of ASCII characters by the fact  the most significant bit of each byte is zero.
Base64 code is made of 64~ASCII characters: all 26~letters (upper case and lower case), all ten digits and two other characters ('\texttt{+}', '\texttt{/}'). Each of these 64~characters correspond to a 6-bit unsigned integer value in the interval [0, 64).
Table~\ref{tab:base64table} provides  
a bidirectional mapping between the integer values and ASCII characters.
 This map is  invertible.

Given three arbitrary byte values spanning $3 \times 8 =24$~bits, we can always represent them using four ASCII characters from the base64 set since $4 \times 6 = 24$~bits. 
During encoding, if the input is not divisible by three bytes, 
then one or two special 
padding character ('\texttt{=}') may be appended so that the length of the
base64 code is divisible into blocks of four ASCII characters.
Given three byte values $s_1, s_2, s_3$, the base64 standard
maps them bijectively to the four 6-bit values $s_1 \div 4$, 
$(s_2 \div 16) + (s_1 \times 16) \bmod 64$, 
$(s_2 \times 4)\bmod 64 + (s_3 \div 64) $ and
$s_3 \bmod 64$.
 We can also use other lists of ASCII characters as in the  base64url standard\cite{rfc4648} where the characters '\texttt{+}' and '\texttt{/}' are replaced by '\texttt{-}' (minus) and '\texttt{\_}' (underline).

We can quickly encode and decode base64 strings using look-up tables. It is the approach taken by the Google Chrome browser.
For example, we can map ASCII characters to 
integers in [0,64) as in  Table~\ref{tab:base64table}, with a special value to identify ASCII characters outside of the base64 domain (for error detection). Given four bytes mapped to integer values in [0,64), $a,b,c,d$, we can retrieve that original three bytes as
$(a\times 4) + (b \div 16), (b \times 16) \bmod 256 + (c \div 4), (c \times 64) \bmod 256 + d$.

\begin{table}
\caption{Base64 mapping\label{tab:base64table} between 6-bit values and ASCII characters.}%
\centering
\begin{tabular}{ccc|ccc|ccc|ccc}
\toprule
value & ASCII & char & value & ASCII & char & value & ASCII & char & value & ASCII & char \\ 
\midrule
0 & 0x41 & \texttt{A} & 16 & 0x51 & \texttt{Q} & 32 & 0x67 & \texttt{g} & 48 & 0x77 & \texttt{w} \\
1 & 0x42 & \texttt{B} & 17 & 0x52 & \texttt{R} & 33 & 0x68 & \texttt{h} & 49 & 0x78 & \texttt{x} \\
2 & 0x43 & \texttt{C} & 18 & 0x53 & \texttt{S} & 34 & 0x69 & \texttt{i} & 50 & 0x79 & \texttt{y} \\
3 & 0x44 & \texttt{D} & 19 & 0x54 & \texttt{T} & 35 & 0x6a & \texttt{j} & 51 & 0x7a & \texttt{z} \\
4 & 0x45 & \texttt{E} & 20 & 0x55 & \texttt{U} & 36 & 0x6b & \texttt{k} & 52 & 0x30 & \texttt{0} \\
5 & 0x46 & \texttt{F} & 21 & 0x56 & \texttt{V} & 37 & 0x6c & \texttt{l} & 53 & 0x31 & \texttt{1} \\
6 & 0x47 & \texttt{G} & 22 & 0x57 & \texttt{W} & 38 & 0x6d & \texttt{m} & 54 & 0x32 & \texttt{2} \\
7 & 0x48 & \texttt{H} & 23 & 0x58 & \texttt{X} & 39 & 0x6e & \texttt{n} & 55 & 0x33 & \texttt{3} \\
8 & 0x49 & \texttt{I} & 24 & 0x59 & \texttt{Y} & 40 & 0x6f & \texttt{o} & 56 & 0x34 & \texttt{4} \\
9 & 0x4a & \texttt{J} & 25 & 0x5a & \texttt{Z} & 41 & 0x70 & \texttt{p} & 57 & 0x35 & \texttt{5} \\
10 & 0x4b & \texttt{K} & 26 & 0x61 & \texttt{a} & 42 & 0x71 & \texttt{q} & 58 & 0x36 & \texttt{6} \\
11 & 0x4c & \texttt{L} & 27 & 0x62 & \texttt{b} & 43 & 0x72 & \texttt{r} & 59 & 0x37 & \texttt{7} \\
12 & 0x4d & \texttt{M} & 28 & 0x63 & \texttt{c} & 44 & 0x73 & \texttt{s} & 60 & 0x38 & \texttt{8} \\
13 & 0x4e & \texttt{N} & 29 & 0x64 & \texttt{d} & 45 & 0x74 & \texttt{t} & 61 & 0x39 & \texttt{9} \\
14 & 0x4f & \texttt{O} & 30 & 0x65 & \texttt{e} & 46 & 0x75 & \texttt{u} & 62 & 0x2b & \texttt{+} \\
15 & 0x50 & \texttt{P} & 31 & 0x66 & \texttt{f} & 47 & 0x76 & \texttt{v} & 63 & 0x2f & \texttt{/} \\
\bottomrule
\end{tabular}
\end{table}
\section{Algorithmic Design}

We seek to encode to base64 and decode from base64 using as few instructions as possible. To this end, we load our data into 512-bit (64~bytes) registers and we use instructions that operate on full 512-bit registers. Because encoding \emph{expands} the number of bytes, we need to consume 48~bytes if we want the encoder to produce 64~bytes with each iteration. Similarly, we want the decoder to consume 64~bytes and produce 48~bytes per iteration.

The AVX-512 instruction sets\cite{Reinders2013} were first introduced by Intel in 2013. It consists of several distinct instructions sets, not all of which are included on all processors with AVX-512 support. 
Within AVX-512, we use instructions from the \emph{Vector Byte Manipulation Instructions} (VBMI) set which we expect to be included in   all new Intel microarchitectures (e.g., Cannon Lake, Cascade Lake, Cooper Lake, Ice Lake) for laptops and servers. 
AVX-512 instructions have been used to accelerate the implementation of algorithms in genomics\cite{Rucci2019}, machine learning\cite{BNCSS117}, databases\cite{Lasch:2019:FSC:3329785.3329924,Zarubin:2019:ICN:3329785.3329923,kersten18} as well as in high-performance computing.

\subsection{Encoding}

If we omit load and store instructions, only two \texttt{vpermb} and
one \texttt{vpmultishiftqb}, i.e.\ just three instructions, 
are necessary
to encode 48~bytes into base64.  In contrast, the previously best
vectorized base64 encoder\cite{simdbase64} based on AVX2 and 256-bit registers
uses 11~SIMD instructions to encode 24~bytes into base64.
In other words, we achieve a seven-fold reduction in instruction count for a given
number of bytes---it exceeds the reduction by a factor of two that we may expect from a doubling of the register size  from 256 to 512~bits.

We use only two distinct AVX-512 instructions:
\begin{itemize}
\item The \texttt{vpermb} instruction takes two 512-bit registers as inputs and treats them as arrays
of 64~bytes. One register is  an array of indexes ($x$), the other  is a collection
of values ($a$). 
The instruction computes the composition of the values and the indexes, $(a[x[0]], a[x[1]], \ldots, a[x[63]])$,
with the convention that only the six least significant digits of index values  are used, the two most significant bits are silently ignored.

The characteristics of this instruction are particularly useful because base64 has exactly 64~ASCII output characters.
\item The  \texttt{vpmultishiftqb} instructions takes two 512-bit registers as inputs. One register is made of eight 64-bit words,
whereas the other register is an array of 64~bytes that are interpreted as shifts. The instruction executes a
total of 64~shifts.

The instruction processes each 64-bit word, separately. Each such 64-bit word from the first register is paired with eight shift bytes from the second register.

Given a 64-bit word and a given shift byte, we take the least significant 6~bits of the
shift byte (as a number between 0 and 64), rotate right the  64-bit word by this amount,
and then output the least significant 8~bits.
\end{itemize}

We can separate the encoding task into two steps. Our goal is to map each sequence of three bytes
 $s_1, s_2, s_3$ to the four 6-bit values $s_1 \div 4$, 
$(s_2 \div 16) + (s_1 \times 16) \bmod 64$, 
$(s_2 \times 4)\bmod 64 + (s_3 \div 64) $ and
$s_3 \bmod 64$.

\begin{enumerate}
    \item We first want to map each block of three 8-bit values to a block of four 6-bit values. 
We plan to produce 64~ASCII characters so we consume only first 48~bytes of the register, ignoring the last 16~bytes of the 64-byte register.
      
    We use the \texttt{vpermb} instruction to map  each sequence of three bytes
    ($s1,s2,s3$) into a sequence of four bytes ($s2,s1,s3,s2$).
Thus if we start with bytes values 0 to 63 (grouped in sets of three)
\begin{center}
\begin{tabular}{ccc|ccc}
    0& 1& 2&   3& 4& 5\\ 
    6& 7& 8&   9& 10& 11\\
    12& 13& 14&   15& 16& 17\\ 
    18& 19& 20&   21& 22& 23\\ 
    24& 25& 26&   27& 28& 29\\ 
    30& 31& 32&   33& 34& 35\\ 
    36& 37& 38&   39& 40& 41\\ 
    42& 43& 44&   45& 46& 47\\ 
    48& 49& 50&   51& 52& 53\\ 
    54& 55& 56&   57& 58& 59\\ 
    60& 61& 62&   63, &  & \\
\end{tabular}\end{center}
we would finish, after the mapping, with byte values (grouped in sets of four)\begin{center}
\begin{tabular}{cccc|cccc}
    1& 0& 2& 1& 4& 3& 5& 4\\ 
7& 6& 8& 7& 10& 9& 11& 10\\
13& 12& 14& 13& 16& 15& 17& 16\\
19& 18& 20& 19& 22& 21& 23& 22\\ 
25& 24& 26& 25& 28& 27& 29& 28\\  
31& 30& 32& 31& 34& 33& 35& 34\\  
37& 36& 38& 37& 40& 39& 41& 40\\  
43& 42& 44& 43& 46& 45& 47& 46. \\
\end{tabular}\end{center}
As our example illustrates, the last sixteen values from the input are unused. Moreover, each block of three input bytes occupies 32~bits (4~bytes) in the shuffled register. To complete the encoding, we need to reorganize the bits in each 32-bit block to fit the base64 format.

To arrive at this desired bit layer, we apply the \texttt{vpmultishiftqb} instruction. Each 8-byte word is shifted eight times with the shift values\begin{center}
\begin{tabular}{llll}
10,& 4,& 22,& 16,\\  
10 + 32,& 4 + 32,& 22 + 32,& 16 + 32.\\
\end{tabular}\end{center}
It is best to view our application of the  \texttt{vpmultishiftqb} on an 4-byte basis even though it operates on 8-byte words.
To see what these shifts do, 
let us consider any four packed bytes $s2,s1,s3,s2$. We label the bit values from  $s1$ to a, $s2$ to b, and $s3$ to c. Similarly add another sequence of four bytes, this time with the labels d,e,f. We start with the following content 
\begin{verbatim}
[bbbbbbbB|aaaaaaaA|cccccccC|bbbbbbbB|eeeeeeeE|dddddddD|fffffffF|eeeeeeeE].
\end{verbatim}
We use the upper case letter to indicate the most significant bit of each byte value.
We then get the following after applying the various right shifts (10, 4, \ldots) and selecting only the least significant eight bits:
\begin{verbatim}
10   : aaaaaAcc,
4    : bbbBaaaa,
22   : cCbbbbbb,
16   : cccccccC,
10+32: dddddDff,
4+32 : eeeEdddd,
22+32: fFeeeeee,
16+32: fffffffF.
\end{verbatim}
We can then pack the result to determine the result of the  \texttt{vpmultishiftqb} instruction:
\begin{verbatim}
[aaaaaAcc|bbbBaaaa|cCbbbbbb|cccccccC|dddddDff|eeeEdddd|fFeeeeee|fffffffF].
\end{verbatim}    
Let us ignore the most significant two bits of each byte:
\begin{verbatim}
[aaaaaA--|bbbBaa--|cCbbbb--|cccccc--|dddddD--|eeeEdd--|fFeeee--|ffffff--].
\end{verbatim}    
If we ignore the two most significant bits of each byte, we see that starting from sequences of four byte values $s2,s1,s3,s2$, we output the four byte values $s_1 \div 4$, 
$(s_2 \div 16) + (s_1 \times 16) \bmod 64$, 
$(s_2 \times 4)\bmod 64 + (s_3 \div 64) $ and
$s_3 \bmod 64$. We do not have to explicitly reset these two bits because, in the upcoming second step, the \texttt{vpermb} instruction does it internally.

 The \texttt{vpmultishiftqb} allows us to use fewer instructions. In our AVX2 algorithm\cite{simdbase64}, this bit shifting requires as much as five instructions: two bitwise AND instructions, two variable shifts instructions and one bitwise OR instruction.
 
    \item We then need to map the blocks of 6-bit values using the \texttt{vpermb} instruction. The 6-bit values are used as indexes while the values are  the list of possible ASCII values as a 64-byte register:\\ \texttt{ABCDEFGHIJKLMNOPQRSTUVWXYZabcdefghijklmnopqrstuvwxyz0123456789+/}.\\
    We can merely replace a few characters in this register to support other base64 variants\cite{rfc4648} like base64url: in fact any 64-byte mapping is feasible, even if determined dynamically at runtime.
\end{enumerate}

\begin{figure}
    \centering
\input{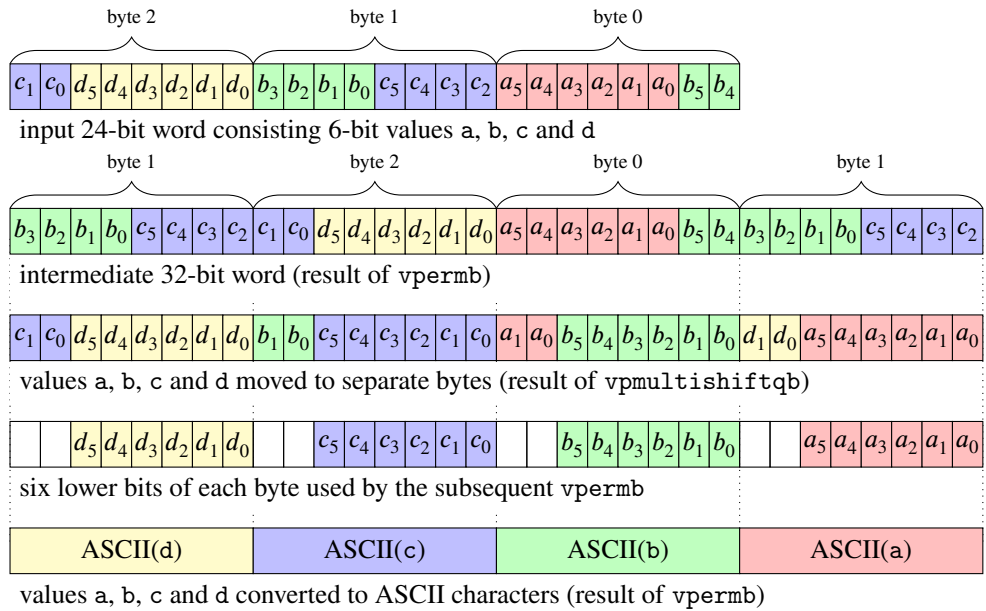}
    \caption{Vectorized encoding algorithm\label{fig:encoding}}
\end{figure}

We summarize the encoding process in Fig.~\ref{fig:encoding}.
In practice, the input to be encoded may not be divisible by 48~bytes: we process any leftover bytes using a conventional code path.

\subsection{Decoding}

One might expect decoding to be just like encoding but in reverse. However, we must validate that the input is proper base64 text. Though the encoding requires only three instructions per block of 64~bytes, we need five instructions to decode 64~bytes: \texttt{vpermi2b}, \texttt{vpternlogd},
\texttt{vpmaddubsw}, \texttt{vpmaddwd}, and \texttt{vpermb}. A single call to an additional instruction (\texttt{vpmovb2m}) is needed, once per base64 stream.
In contrast, the best 256-bit vectorized decoder\cite{simdbase64} uses
14~SIMD instructions to decode 32 input ASCII bytes into 24 output bytes. 
Thus we almost achieve a five-fold reduction in the instruction count for the same number of input bytes. Again, it is far better than the reduction by a factor of two that the wider registers would suggest.

In addition to the \texttt{vpermb} instruction, already present in the encoder, we use the following five AVX-512 instructions:
\begin{itemize}
\item The \texttt{vpermi2b} instruction is a more powerful version of the \texttt{vpermb} instruction. It takes three 512-bit registers.
   One of them ($x$) acts as an array of 64~byte-valued indexes, only the least significant 7~bits of each byte is considered, the most significant bit is ignored. Thus the register is treated as an array of integer indexes in the interval $[0, 128)$.
 The most significant bit in each index value of $x$ is just ignored. 
  The other two registers ($a$, $b$) form a 128-byte array used for lookup. When the index value is less than 64, one register is used, otherwise the other is used.
  Let $\mathrm{choose}(i,x, a, b)$ output 
$ a[x[i]]$ when $x[i]<64$ and $b[x[i]-64]$, with the convention that $x[i]$ is a 6-bit value in $[0,128)$.
We can formally represent the result of the \texttt{vpermi2b} instruction as
$(\mathrm{choose}(0,x, a, b), \mathrm{choose}(1,x, a, b), \ldots, \mathrm{choose}(64,x, a, b))$. 
\item The \texttt{vpmovb2m} instruction creates a 64-bit mask value made of all of the most significant bits of all 64 input bytes taken from a single 512-bit register. 
Thus to check that all byte values in a registers are 7-bit values---i.e.\ are smaller than 0x80---it suffices to check whether the result of the \texttt{vpmovb2m} instruction is zero.

\item The \texttt{vpternlogd}  computes any ternary Boolean function (e.g., A OR B OR C). The function is selected by a programmer by passing an immediate 8-bit integer value to the instruction. This instruction can replace several logical instructions, thus decreasing the instruction count.

\item The multiply-add \texttt{vpmaddubsw} instruction works over pairs of registers which are considered as arrays of thirty-two 16-bit block. 
Each 16-bit block is treated as a vector of two 8-bit integer values. Values from one register are treated as signed bytes $[-128, +127)$ and values from the other register as unsigned bytes $[0, 256)$.

We compute the scalar product of aligned pairs of 16-bit blocks: if the 8-bit integers $x_1$ and $x_2$ come from the first register, and the 8-bit integers $y_1$ and $y_2$ come from the second register, we output $x_1 \times y_1 + x_2 \times y_2$ as a 16-bit value.

\item The \texttt{vpmaddwd} instruction works similarly to the \texttt{vpmaddubsw} instructions but uses 32-bit blocks and multiplying 16-bit signed integers instead.
\end{itemize}

\begin{figure}
    \centering
\begin{tikzpicture}
\node [below, right, fill=white] at (0.00, 0.00) {input --- four bytes};
\draw [fill=red] (0.00, 0.30) rectangle (0.40, 0.90);
\node at (0.20, 0.60) {$1$};
\draw [thin] (0.40, 0.30) rectangle (0.80, 0.90);
\node at (0.60, 0.60) {$1$};
\draw [thin] (0.80, 0.30) rectangle (1.20, 0.90);
\node at (1.00, 0.60) {$0$};
\draw [thin] (1.20, 0.30) rectangle (1.60, 0.90);
\node at (1.40, 0.60) {$0$};
\draw [thin] (1.60, 0.30) rectangle (2.00, 0.90);
\node at (1.80, 0.60) {$0$};
\draw [thin] (2.00, 0.30) rectangle (2.40, 0.90);
\node at (2.20, 0.60) {$0$};
\draw [thin] (2.40, 0.30) rectangle (2.80, 0.90);
\node at (2.60, 0.60) {$1$};
\draw [thin] (2.80, 0.30) rectangle (3.20, 0.90);
\node at (3.00, 0.60) {$1$};
\draw [fill=red] (3.20, 0.30) rectangle (3.60, 0.90);
\node at (3.40, 0.60) {$1$};
\draw [thin] (3.60, 0.30) rectangle (4.00, 0.90);
\node at (3.80, 0.60) {$1$};
\draw [thin] (4.00, 0.30) rectangle (4.40, 0.90);
\node at (4.20, 0.60) {$1$};
\draw [thin] (4.40, 0.30) rectangle (4.80, 0.90);
\node at (4.60, 0.60) {$1$};
\draw [thin] (4.80, 0.30) rectangle (5.20, 0.90);
\node at (5.00, 0.60) {$1$};
\draw [thin] (5.20, 0.30) rectangle (5.60, 0.90);
\node at (5.40, 0.60) {$1$};
\draw [thin] (5.60, 0.30) rectangle (6.00, 0.90);
\node at (5.80, 0.60) {$1$};
\draw [thin] (6.00, 0.30) rectangle (6.40, 0.90);
\node at (6.20, 0.60) {$1$};
\draw [thin] (6.40, 0.30) rectangle (6.80, 0.90);
\node at (6.60, 0.60) {$0$};
\draw [thin] (6.80, 0.30) rectangle (7.20, 0.90);
\node at (7.00, 0.60) {$0$};
\draw [thin] (7.20, 0.30) rectangle (7.60, 0.90);
\node at (7.40, 0.60) {$1$};
\draw [thin] (7.60, 0.30) rectangle (8.00, 0.90);
\node at (7.80, 0.60) {$1$};
\draw [thin] (8.00, 0.30) rectangle (8.40, 0.90);
\node at (8.20, 0.60) {$0$};
\draw [thin] (8.40, 0.30) rectangle (8.80, 0.90);
\node at (8.60, 0.60) {$1$};
\draw [thin] (8.80, 0.30) rectangle (9.20, 0.90);
\node at (9.00, 0.60) {$1$};
\draw [thin] (9.20, 0.30) rectangle (9.60, 0.90);
\node at (9.40, 0.60) {$0$};
\draw [thin] (9.60, 0.30) rectangle (10.00, 0.90);
\node at (9.80, 0.60) {$0$};
\draw [thin] (10.00, 0.30) rectangle (10.40, 0.90);
\node at (10.20, 0.60) {$1$};
\draw [thin] (10.40, 0.30) rectangle (10.80, 0.90);
\node at (10.60, 0.60) {$1$};
\draw [thin] (10.80, 0.30) rectangle (11.20, 0.90);
\node at (11.00, 0.60) {$0$};
\draw [thin] (11.20, 0.30) rectangle (11.60, 0.90);
\node at (11.40, 0.60) {$0$};
\draw [thin] (11.60, 0.30) rectangle (12.00, 0.90);
\node at (11.80, 0.60) {$1$};
\draw [thin] (12.00, 0.30) rectangle (12.40, 0.90);
\node at (12.20, 0.60) {$0$};
\draw [thin] (12.40, 0.30) rectangle (12.80, 0.90);
\node at (12.60, 0.60) {$0$};
\draw [decorate,decoration={brace,amplitude=10pt}] (0.00, 0.90) -- (3.20, 0.90) node [midway,above,yshift=10pt] {\footnotesize \texttt{0xc3} (byte 3)};
\draw [decorate,decoration={brace,amplitude=10pt}] (3.20, 0.90) -- (6.40, 0.90) node [midway,above,yshift=10pt] {\footnotesize \texttt{0xff} (byte 2)};
\draw [decorate,decoration={brace,amplitude=10pt}] (6.40, 0.90) -- (9.60, 0.90) node [midway,above,yshift=10pt] {\footnotesize ASCII "6" (byte 1)};
\draw [decorate,decoration={brace,amplitude=10pt}] (9.60, 0.90) -- (12.80, 0.90) node [midway,above,yshift=10pt] {\footnotesize ASCII "d" (byte 0)};
\node [below, right, fill=white] at (0.00, -1.90) {translated input (result of \texttt{vpermi2b})};
\draw [decorate,decoration={brace,amplitude=10pt}] (0.80, -1.00) -- (3.20, -1.00) node [midway,above,yshift=10pt] {\footnotesize field \texttt{d} = \texttt{0x02}};
\draw [decorate,decoration={brace,amplitude=10pt}] (4.00, -1.00) -- (6.40, -1.00) node [midway,above,yshift=10pt] {\footnotesize field \texttt{c} = \texttt{0x00}};
\draw [decorate,decoration={brace,amplitude=10pt}] (7.20, -1.00) -- (9.60, -1.00) node [midway,above,yshift=10pt] {\footnotesize field \texttt{b} = \texttt{0x3a}};
\draw [decorate,decoration={brace,amplitude=10pt}] (10.40, -1.00) -- (12.80, -1.00) node [midway,above,yshift=10pt] {\footnotesize field \texttt{a} = \texttt{0x1d}};
\draw [thin] (0.00, -1.60) rectangle (0.40, -1.00);
\node at (0.20, -1.30) {$0$};
\draw [thin] (0.40, -1.60) rectangle (0.80, -1.00);
\node at (0.60, -1.30) {$0$};
\draw [fill=yellow!25] (0.80, -1.60) rectangle (1.20, -1.00);
\node at (1.00, -1.30) {$0$};
\draw [fill=yellow!25] (1.20, -1.60) rectangle (1.60, -1.00);
\node at (1.40, -1.30) {$0$};
\draw [fill=yellow!25] (1.60, -1.60) rectangle (2.00, -1.00);
\node at (1.80, -1.30) {$0$};
\draw [fill=yellow!25] (2.00, -1.60) rectangle (2.40, -1.00);
\node at (2.20, -1.30) {$0$};
\draw [fill=yellow!25] (2.40, -1.60) rectangle (2.80, -1.00);
\node at (2.60, -1.30) {$1$};
\draw [fill=yellow!25] (2.80, -1.60) rectangle (3.20, -1.00);
\node at (3.00, -1.30) {$0$};
\draw [fill=red] (3.20, -1.60) rectangle (3.60, -1.00);
\node at (3.40, -1.30) {$1$};
\draw [thin] (3.60, -1.60) rectangle (4.00, -1.00);
\node at (3.80, -1.30) {$0$};
\draw [fill=blue!25] (4.00, -1.60) rectangle (4.40, -1.00);
\node at (4.20, -1.30) {$0$};
\draw [fill=blue!25] (4.40, -1.60) rectangle (4.80, -1.00);
\node at (4.60, -1.30) {$0$};
\draw [fill=blue!25] (4.80, -1.60) rectangle (5.20, -1.00);
\node at (5.00, -1.30) {$0$};
\draw [fill=blue!25] (5.20, -1.60) rectangle (5.60, -1.00);
\node at (5.40, -1.30) {$0$};
\draw [fill=blue!25] (5.60, -1.60) rectangle (6.00, -1.00);
\node at (5.80, -1.30) {$0$};
\draw [fill=blue!25] (6.00, -1.60) rectangle (6.40, -1.00);
\node at (6.20, -1.30) {$0$};
\draw [thin] (6.40, -1.60) rectangle (6.80, -1.00);
\node at (6.60, -1.30) {$0$};
\draw [thin] (6.80, -1.60) rectangle (7.20, -1.00);
\node at (7.00, -1.30) {$0$};
\draw [fill=green!25] (7.20, -1.60) rectangle (7.60, -1.00);
\node at (7.40, -1.30) {$1$};
\draw [fill=green!25] (7.60, -1.60) rectangle (8.00, -1.00);
\node at (7.80, -1.30) {$1$};
\draw [fill=green!25] (8.00, -1.60) rectangle (8.40, -1.00);
\node at (8.20, -1.30) {$1$};
\draw [fill=green!25] (8.40, -1.60) rectangle (8.80, -1.00);
\node at (8.60, -1.30) {$0$};
\draw [fill=green!25] (8.80, -1.60) rectangle (9.20, -1.00);
\node at (9.00, -1.30) {$1$};
\draw [fill=green!25] (9.20, -1.60) rectangle (9.60, -1.00);
\node at (9.40, -1.30) {$0$};
\draw [thin] (9.60, -1.60) rectangle (10.00, -1.00);
\node at (9.80, -1.30) {$0$};
\draw [thin] (10.00, -1.60) rectangle (10.40, -1.00);
\node at (10.20, -1.30) {$0$};
\draw [fill=red!25] (10.40, -1.60) rectangle (10.80, -1.00);
\node at (10.60, -1.30) {$0$};
\draw [fill=red!25] (10.80, -1.60) rectangle (11.20, -1.00);
\node at (11.00, -1.30) {$1$};
\draw [fill=red!25] (11.20, -1.60) rectangle (11.60, -1.00);
\node at (11.40, -1.30) {$1$};
\draw [fill=red!25] (11.60, -1.60) rectangle (12.00, -1.00);
\node at (11.80, -1.30) {$1$};
\draw [fill=red!25] (12.00, -1.60) rectangle (12.40, -1.00);
\node at (12.20, -1.30) {$0$};
\draw [fill=red!25] (12.40, -1.60) rectangle (12.80, -1.00);
\node at (12.60, -1.30) {$1$};
\node [below, right, fill=white] at (0.00, -3.30) {error = input \texttt{OR} translated input \texttt{OR} previous error (result of \texttt{vpternlogd})};
\draw [fill=red] (0.00, -3.00) rectangle (0.40, -2.40);
\node at (0.20, -2.70) {$1$};
\draw [thin] (0.40, -3.00) rectangle (0.80, -2.40);
\node at (0.60, -2.70) {--};
\draw [thin] (0.80, -3.00) rectangle (1.20, -2.40);
\node at (1.00, -2.70) {--};
\draw [thin] (1.20, -3.00) rectangle (1.60, -2.40);
\node at (1.40, -2.70) {--};
\draw [thin] (1.60, -3.00) rectangle (2.00, -2.40);
\node at (1.80, -2.70) {--};
\draw [thin] (2.00, -3.00) rectangle (2.40, -2.40);
\node at (2.20, -2.70) {--};
\draw [thin] (2.40, -3.00) rectangle (2.80, -2.40);
\node at (2.60, -2.70) {--};
\draw [thin] (2.80, -3.00) rectangle (3.20, -2.40);
\node at (3.00, -2.70) {--};
\draw [fill=red] (3.20, -3.00) rectangle (3.60, -2.40);
\node at (3.40, -2.70) {$1$};
\draw [thin] (3.60, -3.00) rectangle (4.00, -2.40);
\node at (3.80, -2.70) {--};
\draw [thin] (4.00, -3.00) rectangle (4.40, -2.40);
\node at (4.20, -2.70) {--};
\draw [thin] (4.40, -3.00) rectangle (4.80, -2.40);
\node at (4.60, -2.70) {--};
\draw [thin] (4.80, -3.00) rectangle (5.20, -2.40);
\node at (5.00, -2.70) {--};
\draw [thin] (5.20, -3.00) rectangle (5.60, -2.40);
\node at (5.40, -2.70) {--};
\draw [thin] (5.60, -3.00) rectangle (6.00, -2.40);
\node at (5.80, -2.70) {--};
\draw [thin] (6.00, -3.00) rectangle (6.40, -2.40);
\node at (6.20, -2.70) {--};
\draw [fill=gray] (6.40, -3.00) rectangle (6.80, -2.40);
\node at (6.60, -2.70) {$0$};
\draw [thin] (6.80, -3.00) rectangle (7.20, -2.40);
\node at (7.00, -2.70) {--};
\draw [thin] (7.20, -3.00) rectangle (7.60, -2.40);
\node at (7.40, -2.70) {--};
\draw [thin] (7.60, -3.00) rectangle (8.00, -2.40);
\node at (7.80, -2.70) {--};
\draw [thin] (8.00, -3.00) rectangle (8.40, -2.40);
\node at (8.20, -2.70) {--};
\draw [thin] (8.40, -3.00) rectangle (8.80, -2.40);
\node at (8.60, -2.70) {--};
\draw [thin] (8.80, -3.00) rectangle (9.20, -2.40);
\node at (9.00, -2.70) {--};
\draw [thin] (9.20, -3.00) rectangle (9.60, -2.40);
\node at (9.40, -2.70) {--};
\draw [fill=gray] (9.60, -3.00) rectangle (10.00, -2.40);
\node at (9.80, -2.70) {$0$};
\draw [thin] (10.00, -3.00) rectangle (10.40, -2.40);
\node at (10.20, -2.70) {--};
\draw [thin] (10.40, -3.00) rectangle (10.80, -2.40);
\node at (10.60, -2.70) {--};
\draw [thin] (10.80, -3.00) rectangle (11.20, -2.40);
\node at (11.00, -2.70) {--};
\draw [thin] (11.20, -3.00) rectangle (11.60, -2.40);
\node at (11.40, -2.70) {--};
\draw [thin] (11.60, -3.00) rectangle (12.00, -2.40);
\node at (11.80, -2.70) {--};
\draw [thin] (12.00, -3.00) rectangle (12.40, -2.40);
\node at (12.20, -2.70) {--};
\draw [thin] (12.40, -3.00) rectangle (12.80, -2.40);
\node at (12.60, -2.70) {--};
\end{tikzpicture}
    \caption{Decoding from ASCII into 6-bit values with error detection\label{fig:decoding1}}
\end{figure}
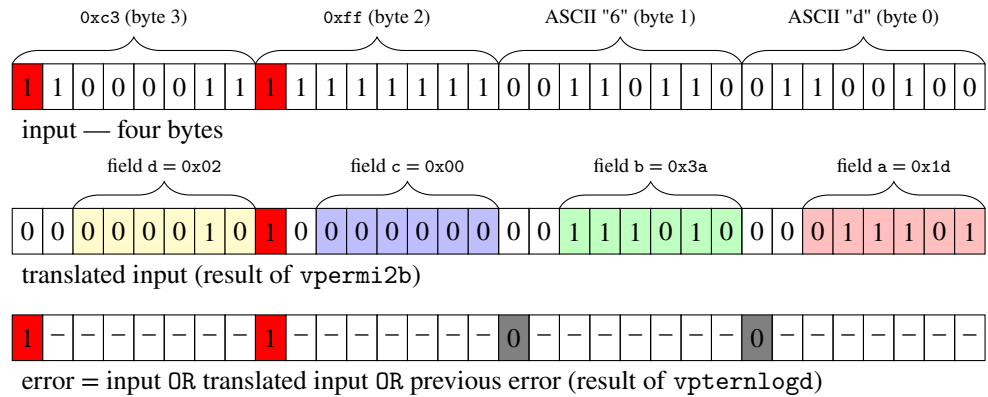

We loop over 512-bit inputs, i.e. 64~ASCII characters.
We first translate the ASCII characters into 6-bit values stored in separate bytes. At this stage, we must detect errors; that is,
we must detect invalid ASCII characters outside of the base64 table.
To translate the ASCII characters to 6-bit values while detecting bad characters, the
 \texttt{vpermi2b} instruction is ideally suited. 
 We set all byte values in the two lookup arrays to byte values 0x80 except for values
 at indexes corresponding to base64 ASCII code points: 0x41, 0x42, \ldots which are set to their
 corresponding 6-bit value, as per Table~\ref{tab:base64table}. 
 If all input values are allowable base64 ASCII characters, we get the correct
 6-bit values stored in a corresponding byte. 
 If any unallowed ASCII character is encountered, then it will result
 in the 0x80 byte value. 
 Furthermore, any non-ASCII code point is characterized as an  input byte
 with the most significant bit set.
 Thus if we take the resulting register  from the \texttt{vpermi2b} instruction and that we
 compute the bitwise OR with the original input, the input contains an incorrect character
 if an only if we can detect a byte value with its most significant bit set.
 If we want to report errors immediately, we can make this check with the \texttt{vpmovb2m} instruction followed by a conditional branch instruction. However, we may postpone error reporting to the end, avoiding branching in the hot loop.
We initialize a 512-bit  \textsc{error} register with zeros, and then repeatedly 
 replace its value with its bitwise OR with the bitwise OR of the input register and of
 the result  \texttt{vpermi2b} instruction. The two bitwise OR operations can 
 be expressed with a single instruction \texttt{vpternlogd}.
 That is, the \texttt{vpmovb2m} instruction followed by a branch is replaced by a single instruction (\texttt{vpternlogd}).
 At the end of the processing, we can check the most significant bit values of the bytes of the
 \textsc{error} register are zero. 
See Fig.~\ref{fig:decoding1}.

The final stage of decoding consists in packing all 6-bit fields into a continuous array of 48 bytes. We proceed in two steps.

\begin{itemize}
    \item  We first need to pack four 6-bit values int into 24-bit words.
    Thus we need to modify, in parallel, the sixteen 32-bit words contained in our working register.
    We achieve this result by two
    multiply-add instructions, operating over 32-bit words.
    We can represent our initial stage within each 32-bit words as follows:
\begin{verbatim}
[00dddddd|00cccccc|00bbbbbb|00aaaaaa].
\end{verbatim}    
If we represent the four byte values as $D, C, B, A$, we want to first replace the sequence of two 8-bit values $D,C$, by the 16-bit value $D + C \times 2^6$; and similarly for the sequence $B,A$. We can get this result with the \texttt{vpmaddubsw} by combining our working register with a constant register made of the values $(0,2^6, 0, 2^6, \ldots)$. The  result on a 32-bit word basis is 
\begin{verbatim}
[0000cccc|ccdddddd|0000aaaa|aabbbbbb].
\end{verbatim}
Finally, we need to combine successive 16-bit words, which we can do in a similar manner, this time using the  \texttt{vpmaddwd} instruction and multiplying our working register with the constant $(2^{12}, 1, 2^{12}, 1, \ldots)$. We get
\begin{verbatim}
[00000000|aaaaaabb|bbbbcccc|ccdddddd].
\end{verbatim}

\item  One byte out of each 4~bytes is zero. We need to \emph{pack} the result into 48~consecutive bytes, in the order specified by the base64 standard. We can achieve this result with the \texttt{vpermb} instruction, using the 48~indexes
\begin{verbatim}
6,0,1,2,9,10,4,5,12,13,14,8,22,16,17,18,25,26,20,21,28,29,30,24,38,32,33,
34,41,42,36,37,44,45,46,40,54, 48,49,50,57,58,52,53,60,61,62,56.
\end{verbatim}
 There are 16~trailing bytes that are unused, so the last 16~indexes are irrelevant.

\end{itemize}

\begin{figure}
    \centering
\input{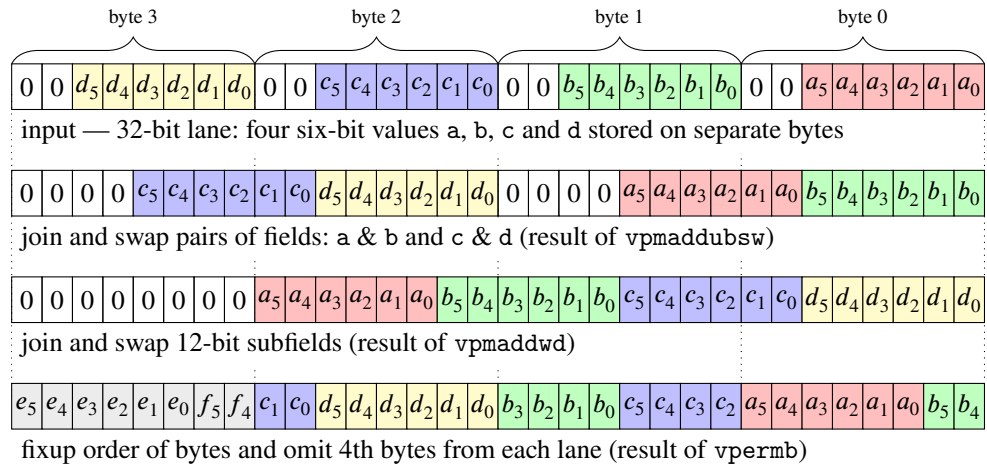}    \caption{Vectorized decoding of three bytes\label{fig:decoding2}}
\end{figure}

Thus the second stage consists in only three instructions:  \texttt{vpmaddubsw}, \texttt{vpmaddwd}, and \texttt{vpermb}.
See Fig.~\ref{fig:decoding2}.

As with the encoding algorithm, we need handle the case where the input is not a multiple of 64~bytes with a final and separate code path.

\section{Experiments}

We summarize the characteristics of our hardware platform in
 Table~\ref{tab:test-cpus}. This processor has 32\,kB of fast (L1) cache per core, 256\,kB of secondary (L2) cache per core and 4\,MB of last-level cache (L3). The system has 
 has a maximal copy (\texttt{memcpy}) bandwidth of $\approx$20\,GB/s.
On some Intel processors, AVX-512 instructions\cite{inteloptimization} are subject to downclocking: whenever AVX-512 instructions are executed, the processor lowers its frequency. We do not observe any downclocking on our processor, nor did we find any Intel documentation referencing downclocking for this processor. To test for downclocking, we used the \texttt{avx-turbo}~\cite{avxturbo} benchmarking tool: it reports the processor frequency after issuing long sequences of expensive instructions of various types, including AVX-512 multiplication and floating-point instructions.

We implemented our software in C. To ease reproducibility, we make it freely available.\footnote{\url{https://github.com/WojciechMula/base64-avx512}}
We compile our code using the GNU~GCC~9.1 compiler with the \texttt{-O3 -march=cannonlake} flags. As a state-of-the-art conventional base64 codec, we use the library used by the Chrome browser. For comparison, we use an optimized AVX2 codec\cite{simdbase64} from our previous work.
To measure the speed, we take 10~measures, compute the median time. Our timings include some fixed overhead costs such as the  function call.

\begin{table*}
\caption{\label{tab:test-cpus} Hardware 
}
\centering
\begin{minipage}{\textwidth}
\centering
\begin{tabular}{cccccc}\toprule
Processor   & Base Frequency & Max. Frequency  & L1 data cache per core & Microarchitecture                          & Compiler\\ \midrule
Intel i3-8121U & 2.2\,GHz  & 3.2\,GHz & 32\,kB & Cannon~Lake (2018) & GCC  9.1 \\
\bottomrule
\end{tabular}
\end{minipage}
\end{table*}

\begin{figure*}\centering
\subfloat[Encoding]{%
\includegraphics[width=0.49\textwidth]{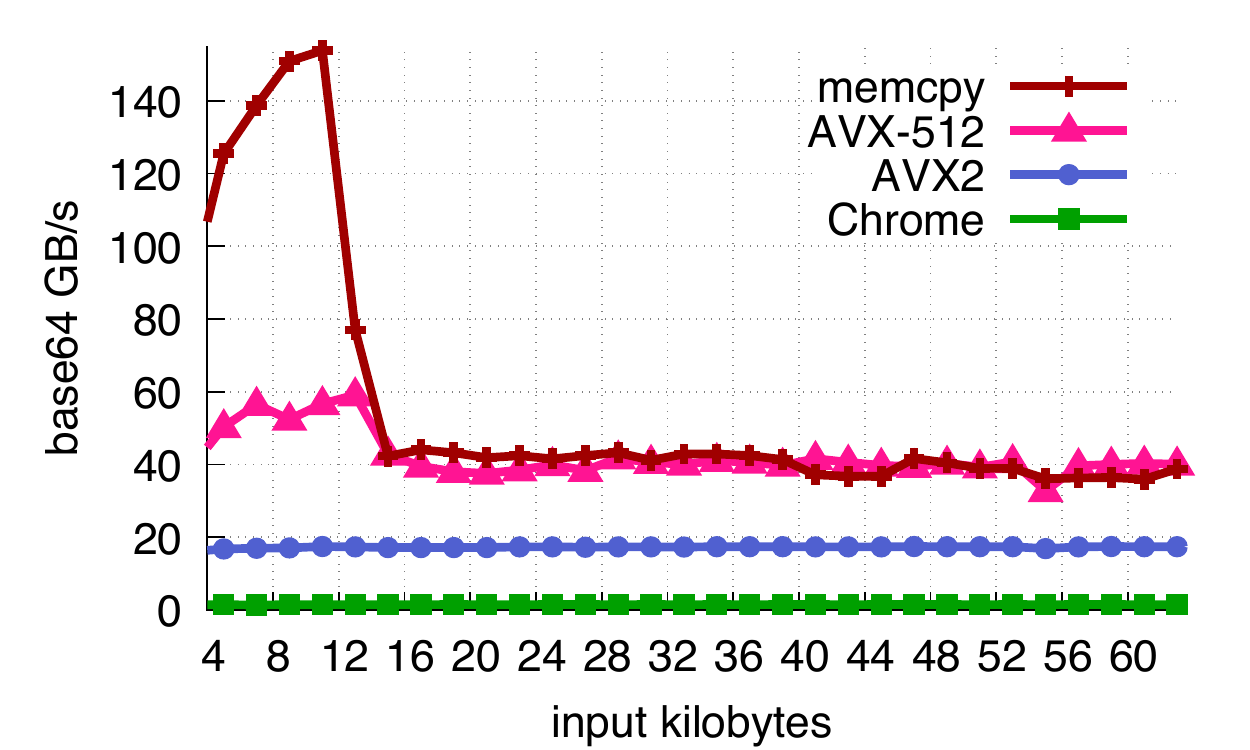}
}\subfloat[Decoding]{%
\includegraphics[width=0.49\textwidth]{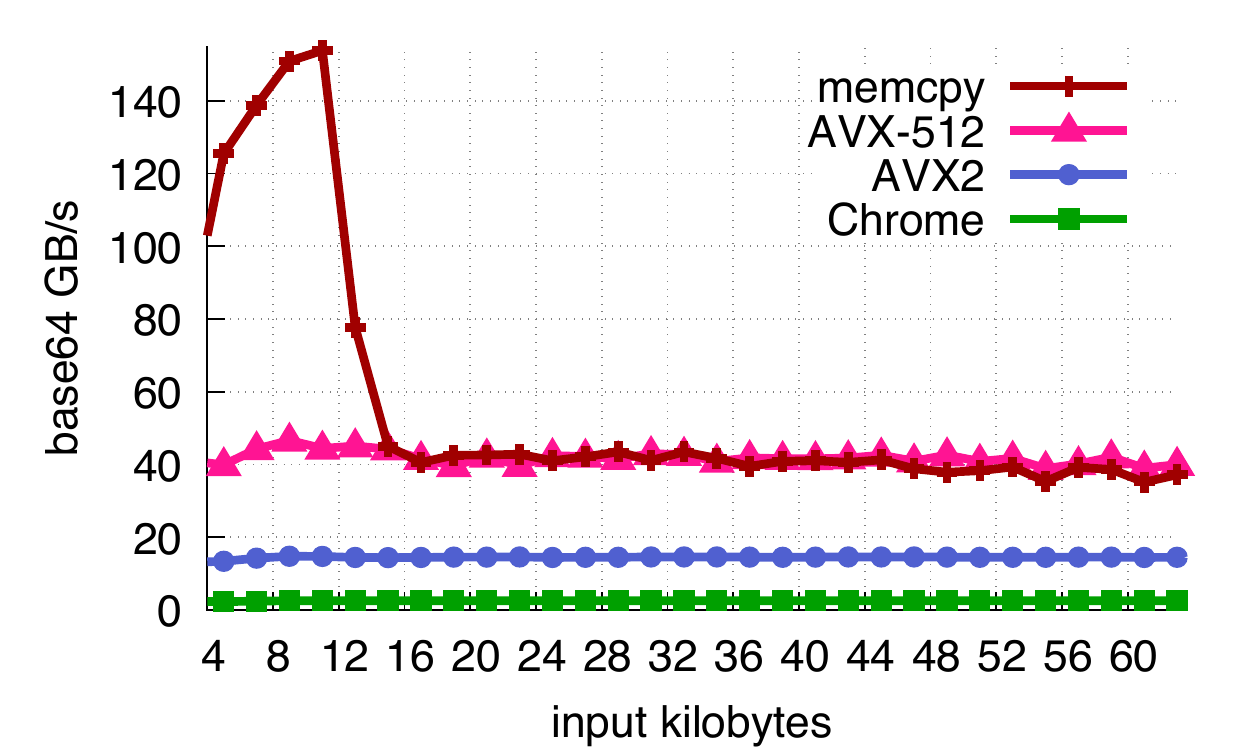}}
\caption{\label{fig:results}Speed in GB/s to encode or decode base64. Data volume is measured in base64 bytes. Speed is lower on tiny inputs due to fixed overheads.}
\end{figure*}

In Fig.~\ref{fig:results}, we present the encoding and decoding speed using random binary data as data source, varying the size from 1\,kB to 64\,kB. Our processor has 32\,kB of L1 cache per core, so we expect that it is able to do a memory copy (\texttt{memcpy}) entirely in L1 cache as long as the input fits in 16\,kB.
We see that for inputs of about 10\,kB, we reach a top copy speed of over 150\,GB/s. This speed is slightly penalized by overhead (function calls and time measurements). 
The speed of the new AVX-512 coded is more than twice that of the state-of-the-art AVX2\cite{simdbase64} codec. This is especially apparent when the data fits in L1 cache (e.g., when the input in less or equal to 12\,kB). 
The speed of the AVX-512 codec is limited to 40\,GB/s for inputs larger than 16\,kB---the same speed also limits the memory copy.
For larger input, the speed of the AVX2 codec is  17\,GB/s for encoding about 15.5\,GB/s for decoding. The conventional codec (Chrome) encodes at 1.5\,GB/s and decodes at 2.6\,GB/s.

We do not expect the vectorized codecs (AVX2 and AVX-512) to be sensitive to the content of the input, keeping the size constant. Nevertheless we run benchmarks on standard images from image processing (Lena, mandril), a Google logo found to be base64 encoded in the Google search page and a large zip file. Our test data is available along with our software. See Table~\ref{tab:perf} for decoding performance.  As expected, the results are similar as with random data. The decoder used in the Chrome browser has a constant speed of 2.6\,GB/s, irrespective of the input size. For the Google logo which fits in L1 cache, the memory copy is faster, but otherwise our AVX-512 is even slightly faster than a memory copy. This is possible because we use the base64 size as a reference when computing the speed: to decode 1\,GB of base64 data, we only need to write about 786\,MB whereas a memory copy must still write a full 1\,GB.  For the large zip file, the AVX-512 is as fast as a memory copy: the benefit of the vectorized decoders (AVX2 and AVX-512)  is less important in this case because we are limited by memory access. In practice, it might be preferable to process large files in small parts that fit in cache when possible to avoid having to write to RAM\@.

\begin{table}%
\caption{Decoding performance in GB/s\label{tab:perf}}%
\centering
\begin{tabular}{|lr|s|sss|}
\toprule
Source & bytes  & \texttt{memcpy} & Chrome &   AVX2 & \multicolumn{1}{c|}{AVX-512} \\\midrule
lena [jpg] &  \num{141020} &\tablenum{25} &  \tablenum{2.6} &  \tablenum{14} & \tablenum{32}  \\
mandril [jpg] & \num{247222}&\tablenum{18} & \tablenum{2.6} & \tablenum{14} & \tablenum{25}\\
Google logo [png] & \num{2357}&\tablenum{44} & \tablenum{2.6} &  \tablenum{14} & \tablenum{42}\\
large [zip] & \num{34904444}&\tablenum{9.5} & \tablenum{2.6} &  \tablenum{8.3} & \tablenum{9.5}\\
\bottomrule
\end{tabular}
\end{table}

\section{Conclusion}


A straight-forward extension of the best base64 codec using 256-bit registers (AVX2) to 512-bit registers (AVX-512) should reduce the instruction count by a factor of two in the best scenario. Yet, not counting load and store instructions, we reduce the instruction count by a factor of seven (for decoding) and five (for encoding). The net result is that we can encode and decode base64 data at a speed of 40\,GB/s when the data is in secondary (L2) cache. For many applications, we expect that encoding and decoding base64 might become a negligible computational burden with our AVX-512 approach: our speed is limited by data access. Our results illustrate the value of powerful SIMD instructions: our codec is 10 to 20~times faster than a highly optimized conventional codec.

The AVX-512 instructions are currently specific to Intel processors. However, wider and more powerful SIMD instructions are already available on some ARM processors\cite{7924233} via the SVE extension, with an even more powerful extension (SVE2) having been announced. We expect that such new instruction sets should be applicable to base64 decoding and encoding.
Future work could also integrate fast base64 decoders inside vectorized parsers such as simdjson~\cite{simdjson}.

 \section*{Acknowledgments}
The work is supported by the
Natural Sciences and Engineering Research Council of Canada
under grant RGPIN-2017-03910.

\bibliographystyle{WileyNJD-AMA}
\bibliography{avx512base64}

\end{document}